\documentclass[aps,prb,showpacs,amsmath,twocolumn,amssymb,superscriptaddress,letterpaper]{revtex4}
\usepackage{times}
\usepackage{amsfonts}
\usepackage{mathrsfs}
\usepackage{graphicx}
\usepackage{dcolumn}
\usepackage{bm}
\usepackage{color}

\usepackage[colorlinks,bookmarks=false,citecolor=blue,linkcolor=red,urlcolor=blue]{hyperref}
\def\be{\begin{equation}}       \def\ee{\end{equation}}
\def\bea{\begin{eqnarray}}      \def\eea{\end{eqnarray}}

\begin{document}
\title{Magnetic ordering and multiferroicity in $\textrm{MnI}_2$ }
\author{Xianxin Wu}
\affiliation{ Institute of Physics, Chinese Academy of Sciences,
Beijing 100190, China}
\author{Yingxiang Cai}
\affiliation{ Department of Physics, Nanchang University, Nanchang
330031, China}
\author{Qing Xie}\affiliation{Department of Physics, Ningbo University,
Zhejiang 315211, China} \affiliation{ Institute of Physics, Chinese
Academy of Sciences, Beijing 100190, China}
\author{Hongming Weng} \email{hmweng@iphy.ac.cn}  \affiliation{ Institute of Physics, Chinese Academy of Sciences,
Beijing 100190, China}
\author{Heng Fan } \email{hfan@iphy.ac.cn }  \affiliation{ Institute of Physics, Chinese Academy of Sciences,
Beijing 100190, China}
\author{Jiangping Hu  }\email{jphu@iphy.ac.cn} \affiliation{
Institute of Physics, Chinese Academy of Sciences, Beijing 100190,
China}\affiliation{Department of Physics, Purdue
University, West Lafayette, Indiana 47907, USA}

\date{\today}

\begin{abstract}
Density-functional calculations are carried out  to investigate
incommensurate magnetic structures and ferroelectric polarization in
newly discovered multiferroic material MnI$_2$. The exchange
interactions among local moments on Mn are parameterized by mapping
the mean-field Heisenberg model on to total energy difference of
several magnetic ordering states. The experimentally observed
noncollinear magnetic states are well reproduced by using this model
and exchange interaction parameters. The direction of polarization
experimentally measured is also consistent with the result derived
from the symmetry analysis of the magnetically ordered state.  In
particular, we find that the inter-plane magnetic exchange coupling
is  pivotal to the emergence of  the spiral magnetic structure. This
noncollinear magnetic structure, combined with spin-orbit coupling
mainly from I ions, is responsible for the appearance of
ferroelectric polarization.

\end{abstract}

\pacs{75.85.+t, 75.10.Hk, 71.70.Ej, 71.15.Mb}

\maketitle

\section{Introduction}

Multiferroics, which exhibit magnetic and dielectric order in the
same phase, recently have attracted increasing
attention\cite{Tokura2006,Cheong2007}. The recent experimental
research on multiferroics has shown that ferroelectricity(FE) and
magnetism couple so strongly that the electric degree of freedom can
be manipulated by an external magnetic field and vice
versa\cite{Kimura2003a,Hur2004a,Kigomiya2003,Hur2004b,Chapon2004a,Kadomtseva2006,Katsura2007,Chapon2006,Kimura2007,Choi2008}.
These properties offer unprecedented applications in modern
energy-effective electronic data storage
technology\cite{Auciello1998,Spaldin2010}.

Theoretically, phenomenological models and symmetry analysis have
clarified the circumstances where a spiral spin structure can induce
an electric polarization\cite{Mostovoy2006,Harris2007}.
Harris\cite{Harris2007} gives a simple method to describe the
magnetic ordering  and their relationship to ferroelectricity based
on lattice, space and time reversal symmetries:  the symmetry of the
magnetoelectric interaction can determine the direction of
spontaneous polarization induced by magnetism.

Several microscopic mechanisms have been proposed to explain the
magnetoelectric coupling in multiferroics. One is the well-known
Katsura-Nagaosa-Balatsky (KNB) model\cite{Katsura2005} which is
based on the idea that spin currents are induced between the spiral
spins and can therefore be considered as electric moments. The
second is that the magnetically induced ionic displacements due to
Dzyaloshinskii-Moriya (DM) interactons can lead to
polarization\cite{Sergienko2006,Harris2006}. The electric
cancelation model\cite{Hu2008} gives a simple but general mechanism
to understand the interplay between ferroelectricity and
noncollinear magnetism in multiferroics. As a powerful tool to
investigate the electronic structure of materials, density
functional theory (DFT) has played an important role in the
understanding of the collinear-spin type\cite{Wang2007,Picozzi2007}
and the spiral magnetic materials LiCu$_2$O$_2$ and
LiCuVO$_4$\cite{Xiang2007}.

MnI$_2$ has been investigated primarily due to the interest in
magnetic and optical properties\cite{Cable,Sato,Hoekstra1983}.
However, it has been discovered recently by Kurumaji \emph{et
al}.\cite{Kurumaji} that MnI$_2$ is also a multiferroic material.
MnI$_2$ crystallizes in the CdI$_2$ type structure with the space
group $P\bar{3}m1$ (No.164). The unit cell contains one formula unit
(f.u.) with the manganese ion located at $(0,0,0)$ and the iodide
ions at $\pm(\frac{1}{3},\frac{2}{3},u)$, where $u=0.245\pm0.002$,
$a=4.146$ {\AA} and $c=6.829$ {\AA}\cite{Cable}. Magnetic properties
are dominated by Mn$^{2+}$ ion with $S=\frac{5}{2}$. Sato \emph{et
al}.\cite{Sato} observed three successive phase transitions at 3.95
K $(T_{N1})$, 3.8 K $(T_{N2})$ and 3.45 K $(T_{N3})$. As temperature
decreases, the Bragg reflection at $\textbf{q}_{im}(q_1, q_2,
q_3)\sim(0.1025, 0.1025, 0.5)$ appears at $T_{N1}$. When the
temperature is further decreased, the reflection position begins to
move slightly out of the $(hhl)$ plane towards the $(h0l)$ plane.
Finally at $T_{N3}$ it jumps to $\textbf{q}_{it}\sim(0.181, 0,
0.439)$, in which we notice that $q_1$ is not equal to $q_2$. Below
$T_{N3}$, the proper screw magnetic structure is realized, which
induces FE polarization about 84 $\mu C/m^2$ along [110] direction
at 2 K\cite{Kurumaji}. Moreover, an in-plane external magnetic field
$H$ can induce the rearrangement of the six multiferroic domains and
every 60$^\circ$ rotation of the in-plane $H$ leads to 120$^\circ$
flop of the $P$ direction as a result of the flop of magnetic order.

Important questions concerning MnI$_2$ are why it has the helix spin
magnetic ground state and how the spiral spin induces ferroelectric
polarization. It is also of great interest in the appearance of
successive phase transitions as temperature  decreases. In this
paper we perform a comprehensive theoretical investigation of these
intriguing properties. We first calculate the magnetic exchange
coupling parameters in MnI$_2$ and then discuss the magnetic phase
transitions mentioned above within mean-field theory based on a
Heisenberg-type magnetic exchange Hamiltonian, in which six exchange
interactions are taken into account. And six exchange interactions
are found to be necessary to give a good description of observed
magnetic structure with Heisenberg model. We find that the
inter-plane coupling  is fairly strong because of its linear
exchange path and is extremely important in inducing the spiral
magnetic order ground state. We further calculate the polarization
of MnI$_2$ and perform symmetry analysis to show the polarization is
consistent with magnetic order. Finally, we show that the spin-orbit
coupling (SOC) on I ions makes primary contribution  to FE
polarization, based on an analysis of charge density difference
between cases with and without SOC.

The paper is organized as following. First, in Sec.~\ref{S1}, we
perform DFT calculation to obtain the six exchange parameters from
eight spin ordered arrangements and determine the magnetic
modulation vectors of MnI$_2$ by using these exchange parameters.
Then, in Sec.~\ref{S2}, we determine the direction of the
polarization in MnI$_2$ through symmetry analysis and calculate the
polarization of different magnetic vectors using DFT. Finally, in
Sec.~\ref{S3}, we give a summary and provide the main conclusions of
our paper.

\section{Wave-vector selection in $\textrm{MnI}_2$ }\label{S1}

\subsection{Calculation of the exchange interaction parameters}
\begin{figure}[t]
\centerline{\includegraphics[height=7.0cm]{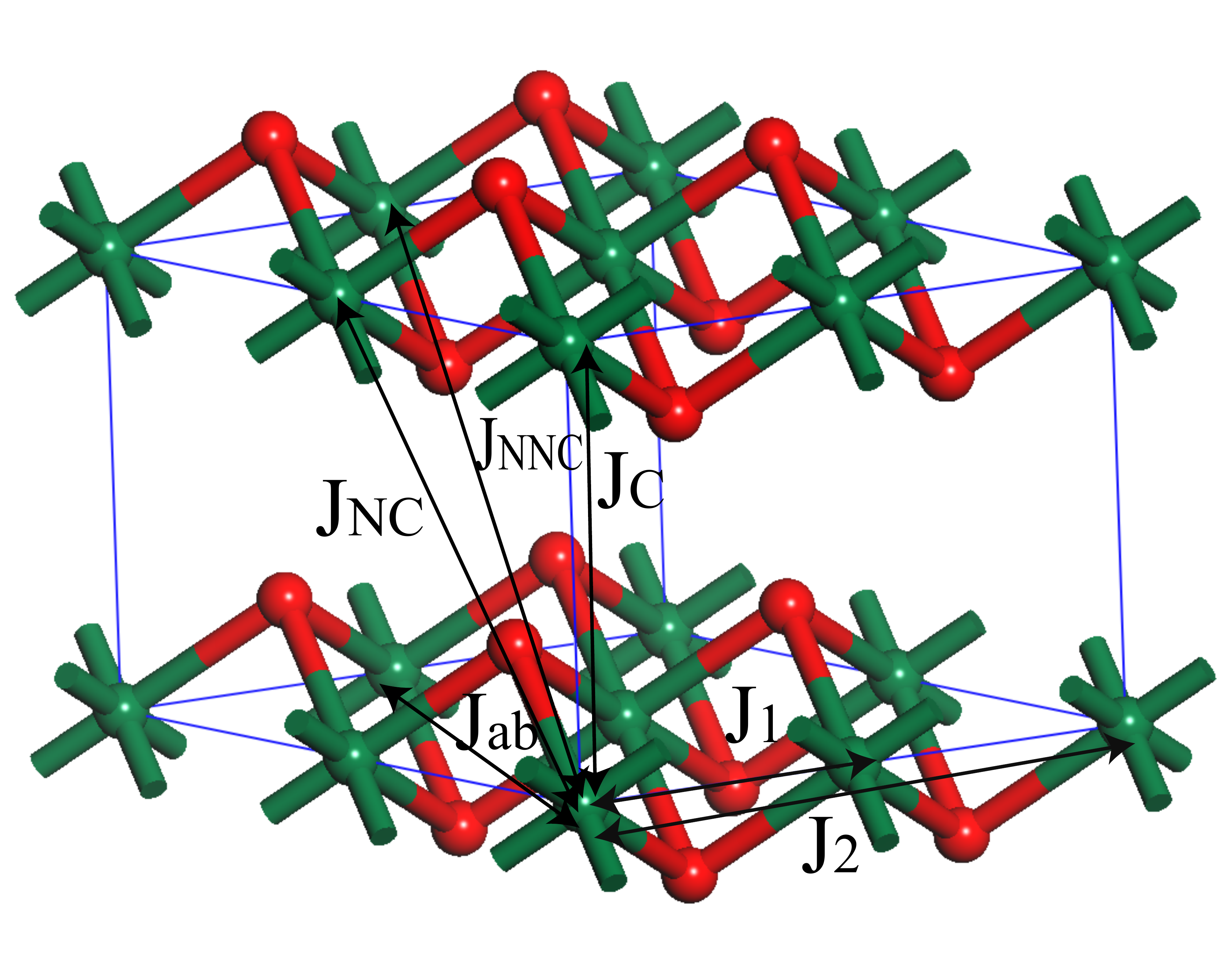}}
\caption{(color online). Side view of the MnI$_2$. The red and green
atoms are Mn and I, respectively. The exchange parameters $J_1$,
$J_{ab}$, $J_2$, $J_C$, $J_{NC}$ and $J_{NNC}$ between the cations
connected by arrows are defined.
 \label{exchange_parameter}}
\end{figure}

 Our DFT calculations employ the
projector augmented wave (PAW) method encoded in Vienna \emph{ab
initio} simulation package(VASP)
\cite{Kresse1993,Kresse1996,Kresse1996B}, and the
generalized-gradient approximation (GGA) for the exchange
correlation functional \cite{Perdew1996} is used. Throughout this
work, the
cutoff energy of 400 eV 
is taken for expanding the wave functions into plane-wave basis. A
set of $2\times4\times2$ $\Gamma$ centered k-points is used for the
$4\times2\times2$ supercell calculation, which is sufficient to
obtain converged results for all quantities under consideration. It
is well-known that GGA underestimates the correlation effect. To
remedy this, the GGA plus on-site repulsion U method (GGA+U) in the
formulation of Dudarev \emph{et al}.\cite{Dudarev} is employed to
describe the electron correlation effect associated with the Mn 3d
states by an effective parameter $U_{eff}$.\cite{Liechtenstein}
Several $U_{eff}$ values for Mn are taken in our calculations to
check the validation of $U_{eff}$. In general, a proper choice of
$U_{eff}$ can systematically reproduce most of the experimental
observations quite well. The self-consistent-field convergence is
achieved when the total electronic energy difference between last
two cycles is less than $10^{-7}$ eV. In all our calculations, we
use the experimental crystal structure\cite{Cable} as shown in
Fig.~\ref{exchange_parameter}. Mn ion is surrounded by octahedron of
I ions and these octahedra are connected by sharing edges to form a
triangle lattice of Mn atom in $ab$ plane stacking along $c$
lattice. Geometrically, it has inversion center and should have no
electrical polarization. As discussed in the following, the
noncollinear magnetic ordering breaks inversion symmetry and induces
experimentally observed polarization, as well as the strong
magneto-electric coupling effects.

In order to obtain the exchange parameters from DFT calculations, we
separate the total energy into nonmagnetic($H_{non}$) and magnetic
contributions

\begin{eqnarray}
H=H_{non}+\sum_{i<j}J_{ij}\hat{S}_i \cdot \hat{S}_j,
\end{eqnarray}
where $\hat{S}_i$ and $\hat{S}_j$ are the spin operators on sites
$i$ and $j$, respectively and the $J_{ij}$ is the exchange
interaction parameter between the sites $i$ and $j$. $J_{ij}<0$
corresponds to the ferromagnetic (FM) coupling between the two sites
while $J_{ij}>0$ corresponds to antiferromagnetic (AFM) interaction.

Fig.~\ref{exchange_parameter} illustrates the magnetic pair exchange
interaction used in our modeling. $J_1$, $J_{ab}$ and $J_2$ are the
intraplane interactions between the cations. $J_C$, $J_{NC}$ and
$J_{NNC}$ are the interplane ones. As it can be seen from
Fig.~\ref{exchange_parameter}, the distance between Mn cations in
the coupling $J_{NNC}$(9.910 \AA) is much longer than that of the
coupling $J_C$(6.829 \AA) and $J_{NC}$(7.989 \AA). One might expect
that the coupling $J_{NNC}$ is much weaker. However, according to
Wollan \emph{et al}.\cite{Wollan} $J_{NNC}$ is fairly strong since
it has an almost linear exchange path
(Mn$^{2+}$-Br$^{-}$-Br$^{-}$-Mn$^{2+}$) in MnBr$_2$, whose structure
is isomorphous with that of MnI$_2$. Our calculation shows that the
magnitude of $J_{NNC}$ is almost the same as $J_C$ and $J_{NC}$,
which confirms their conclusion.

\begin{figure*}
\centerline{\includegraphics[height=7.0cm]{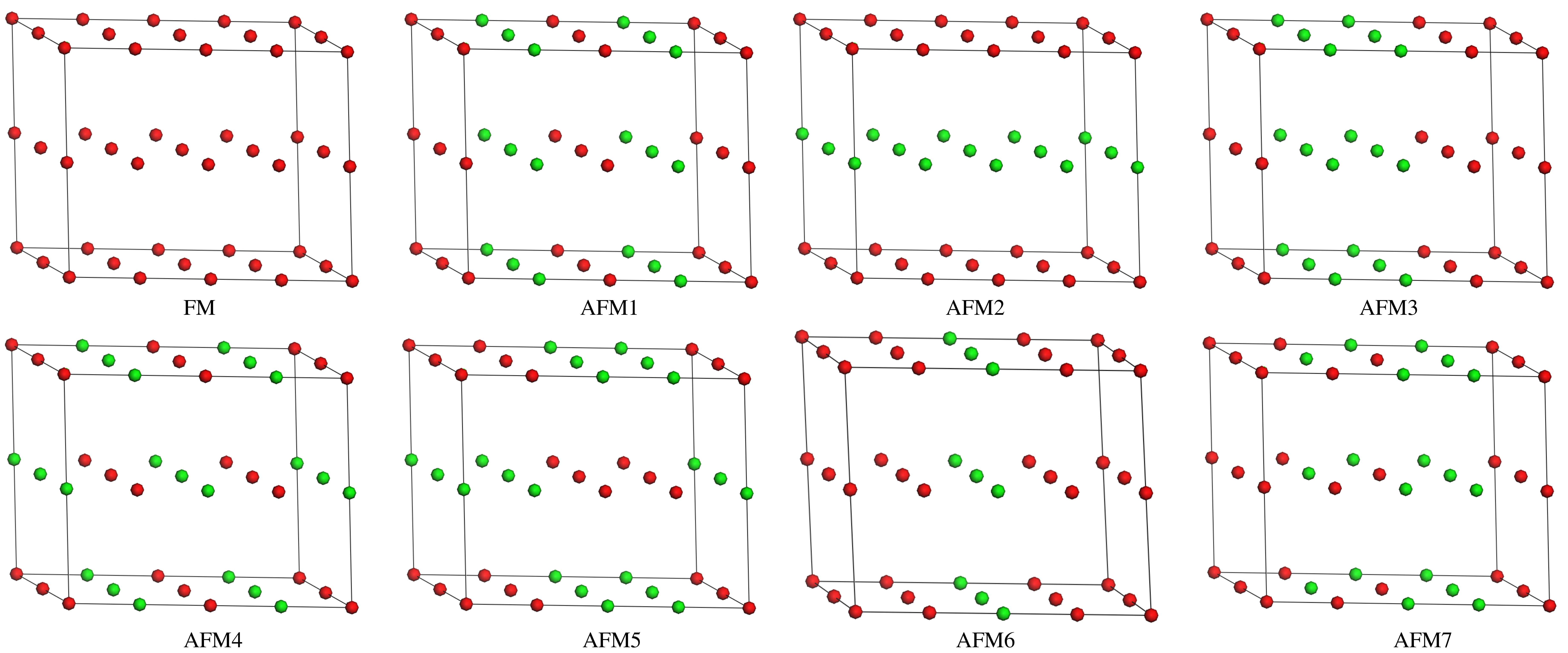}}
\caption{(color online). Schematic plots of eight different magnetic
ordering states of MnI$_2$ used for GGA+U calculation to extract the
six spin-exchange parameters $J_1,J_2,J_{ab},J_C,J_{NC}$ and
$J_{NNC}$. The red and green circles represent the up and down
magnetic moment on Mn sites, respectively.
 \label{states} }
\end{figure*}

  The six spin exchange parameters can be evaluated by examining the
eight ordered spin states of MnI$_2$, i.e. the FM, AFM1, AFM2, AFM3,
AFM4, AFM5, AFM6 and AFM7 states, defined in Fig.~\ref{states} in
terms of a $4 \times 2 \times 2 $ supercell. Table~\ref{Erelative}
summarizes the relative energies of these states per $4 \times 2
\times 2 $ supercell, i.e. 16 f.u., determined from our GGA+U
calculations with and without SOC included. From the energy
expressions obtained for spin dimers with N unpaired spins per site
(in present case, N=5)\cite{Dai}, the energies contributed by
magnetic interactions in these eight magnetic states per f.u. can be
written as

\begin{eqnarray}
E_{FM}&=&\frac{N^2}{4}(3J_1+3J_2+3J_{ab}+J_C+6J_{NC}+3J_{NNC}), \nonumber \\
E_{1}&=&\frac{N^2}{4}(-J_1+3J_2-J_{ab}+J_C-2J_{NC}-J_{NNC}),\nonumber\\
E_{2}&=&\frac{N^2}{4}(3J_1+3J_2+3J_{ab}-J_C-6J_{NC}-3J_{NNC}), \
\notag
\end{eqnarray}
\begin{eqnarray}
E_{3}&=&\frac{N^2}{4}(J_1-J_2-J_{ab}+J_C+2J_{NC}-J_{NNC}),\nonumber\\
E_{4}&=&\frac{N^2}{4}(-J_1+3J_2-J_{ab}-J_C+2J_{NC}+J_{NNC}),\nonumber\\
E_{5}&=&\frac{N^2}{4}(J_1-J_2-J_{ab}-J_C-2J_{NC}+J_{NNC}),\nonumber\\
E_{7}&=&\frac{N^2}{4}(-\frac{1}{2}J_1+J_2-\frac{1}{2}J_{ab}+J_C-\frac{1}{2}J_{NC}+\frac{1}{2}J_{NNC}),\nonumber \\
E_{6}&=&\frac{N^2}{4}(J_1+J_2+J_C+2J_{NC}).
\end{eqnarray}

\begin{table*}[bt]
\caption{\label{Erelative}%
Relative energies (in meV) of eight ordered spin states of MnI$_2$
obtained from the GGA+U calculations with (yes) and without (no) SOC
included for different $U_{eff}$ values. In calculation with SOC,
the spin quantization principle axis is in parallel with $c$ axis. }

\begin{ruledtabular}
\begin{tabular}{cccccccccc}

$U_{eff}(eV)$ & 0 & 0 & 1 & 2 & 3 & 4 & 4 & 5 & 6\\
SOC & no & yes & no & no & no & no & yes & no & no\\
 \colrule

FM & 0 & 0 & 0 & 0 & 0 & 0 & 0 & 0 & 0
\\
$AFM1$ & -258.7 & -250.2 & -149.5 & -81.1 & -35.8 &-4.8 & -2.8 &
16.8 & 32.1
\\
$AFM2$ & -83.1 & -82.9 & -58.4  & -41.5 & -29.6 & -21.0 & -21.0 &
-14.7 & -10.1
\\
$AFM3$ & -203.8 & -199.3 & -127.9 & -79.0 & -45.8 & -22.7 & -21.6 &
-6.3 & 5.4
\\
$AFM4$ & -241.8 & -233.2 & -137.7 & -73.1 & -30.6 &-1.7 & 0.3 & 18.5
& 32.7
\\
$AFM5$ & -210.7 & -206.4 & -133.2  & -83.4 & -49.8 &-26.3 & -25.3 &
-9.6 & 2.4
\\
$AFM6$ & -166.6 & -162.3 & -101.3 & -59.7 & -31.9 & -12.6 & -11.5 &
1.1 & 10.8
\\
$AFM7$ & -253.6 & -245.6 & -150.4 &-85.3 & -41.9 & -12.0 & -10.2 &
9.0 & 23.9
\end{tabular}
\end{ruledtabular}
\end{table*}

 By mapping these onto the total energies obtained from DFT calculations, we obtain
seven equations. But there are only six spin-exchange parameters,
$J_1$, $J_{ab}$, $J_2$ ,$J_C$, $J_{NC}$ and $J_{NNC}$ to be solved.
For this overdetermined system of equations, we obtain these
parameters by using a least-squares
technique\cite{Williams1990,Sims2010}, and list them in
Table~\ref{exchange}.

\begin{table*}[t]
\caption{\label{exchange}%
Values of the spin-exchange parameters J (in meV) in MnI$_2$
obtained from the GGA+U calculations with (yes) and without (no) SOC
included for different $U_{eff}$ values. In calculation with SOC,
the spin quantization principle axis is in parallel with $c$ axis. }
\begin{ruledtabular}
\begin{tabular}{cccccccccc}

$U_{eff}(eV)$ & 0 & 0 & 1 & 2 & 3 & 4 & 4 & 5 & 6\\
SOC & no & yes & no & no & no & no & yes & no & no\\

\colrule $J_1$ & 0.49 & 0.46 & 0.27 & 0.13 & 0.03& -0.03 & -0.03 &
-0.07 & -0.10
\\
$J_2$ & 0.14 & 0.13 & 0.10 & 0.07 & 0.05 & 0.04 & 0.04 & 0.03 & 0.02
\\
$J_{ab}$ & 0.03 & 0.03 & 0.02 & 0.01 & 0.01 & 0.01 & 0.01 & 0 & 0
\\
$J_C$ & 0.04 & 0.04 & 0.03 &0.02 & 0.02 & 0.01 & 0.01 & 0.01 & 0.01
\\
$J_{NC}$ & 0.03 & 0.04 & 0.02 & 0.02 & 0.01 & 0.01 & 0.01 & 0 & 0
\\
$J_{NNC}$ & 0.06 & 0.05 & 0.04 & 0.03 & 0.02 & 0.01 & 0.01 & 0 & 0
\\

\end{tabular}
\end{ruledtabular}
\end{table*}

It is noted from the Table~\ref{exchange} that the intraplane
exchange couplings $J_1$, $J_2$ and $J_{ab}$ are antiferromagnetic
for $U_{eff}\leq 3$ eV. However, $J_1$ becomes negative (i.e. to be
ferromagnetic coupling) with $U_{eff}\geq4$ eV, which is consistent
with the estimated coupling in a similar compound
MnBr$_2$\cite{Sato_MnBr2} from neutron diffraction experiment. As a
consequence, the intraplane spin-exchange interactions are
geometrically frustrated. We  notice that the exchange coupling is
rather weak compared with similar compound such as CuCl$_2$
(exchange parameter is about 10 meV)\cite{Banks}. The weak exchange
coupling can be expected from the observed low magnetic phase
transition temperature (3.45K - 3.95K).  The intraplane exchange
coupling $J_1$ arises from two competing contributions, FM direct
exchange and AFM superexchange interactions between two nearest Mns.
The AFM superexchange interaction is mediated by two Mn-I-Mn bonds
with the same bond angle 90.44$^\circ $. For the similar case of
Cu-O-Cu bonds, it has been shown that when the bond angle is close
to 90$^\circ$ the resulting exchange energy is
rather small.\cite{Tornow} 
For the case of $U_{eff}=4$ eV, one may notice that the coupling
$J_1$ becomes ferromagnetic and weaker than $J_2$ in magnitude,
although the distance of Mn-Mn in $J_2$ coupling is two times of
that in $J_1$. This seems strange but can be easily understood since
both direct exchange and superexchange coupling contribute to $J_1$.
They have opposite signs and compete against each other. As U
increases, the superexchange coupling becomes weaker while the
direct exchange coupling is almost unchanged, which finally leads to
a weak ferromagnetic coupling. $J_1$ becomes dominant for $U_{eff}$=
5 and 6 eV.

The interplane coupling parameters $J_C$, $J_{NC}$ and $J_{NNC}$ are
AFM for all $U_{eff}$, which is consistent with the experiment
carried out by Cable \emph{et al.}\cite{Cable,Sato}. The fact that
this two-anion indirect exchange coupling $J_{NNC}$ appears to be
antiferromagnetic might be expected by analogy with the single-anion
superexchange mechanism. This coupling, which has linear exchange
path (Mn$^{2+}$-I$^{-}$-I$^{-}$-Mn$^{2+}$), is stronger than the
other two interplane couplings for $U_{eff}\leq$ 3 eV. The magnitude
of the three interplane interactions are almost the same for
$U_{eff}=$ 4 eV. $J_{NC}$ and $J_{NNC}$ become zero ($<$ 0.01 meV)
for $U_{eff}$= 5 and 6 eV.

From the theory of superexchange it follows that the corresponding
coupling strength is proportional to $1/U$. If a coupling is
mediated mainly by the superexchange interacting, one will expect a
strong influence of the Hubbard parameter on the strength of this
coupling. That is the reason why most of the exchange coupling
strength decrease significantly with the increasing $U_{eff}$. The
variation of $J_{NNC}$ seems to show a $1/U$ dependence, as being
mediated by superexchange interaction, which is consistent with the
exchange path analysis above.  While $J_{ab}$ has less $U_{eff}$
dependence, which is mainly a direct exchange coupling. We have
further checked that SOC has little influence on exchange
interactions, as shown for GGA and GGA+U ($U_{eff}$=4.0 eV) cases in
Table~\ref{exchange}. Therefore, in the following we use the values
obtained from the calculation without SOC.

\subsection{The classical ground state of the MnI$_2$ } \label{Classical}

\begin{figure}
\centerline{\includegraphics[height=4.0cm]{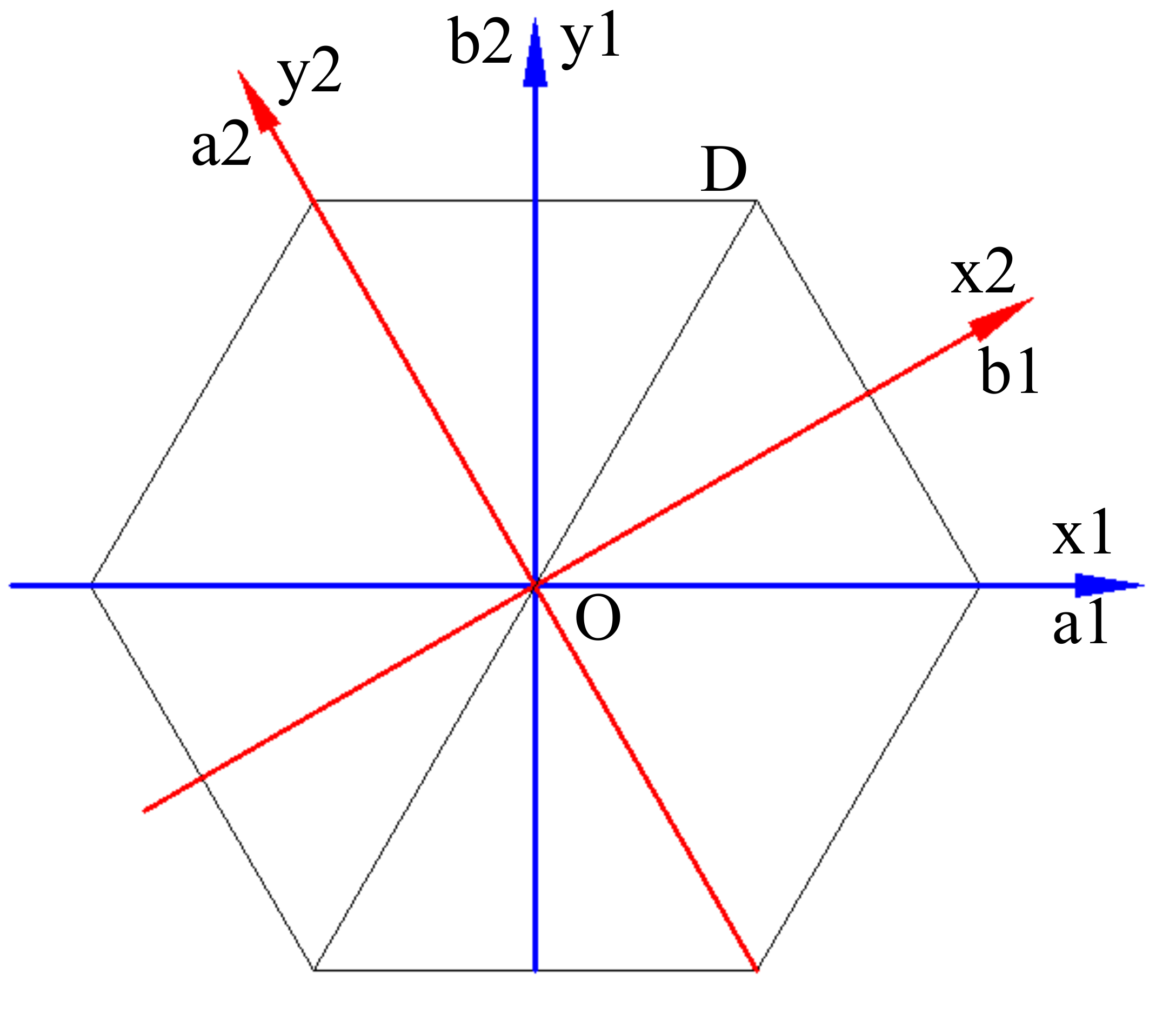}}
 \caption{(color online). The coordinate system in the
paper. $a1$ and $a2$ denote the lattice basis while $b1$ and $b2$
denote the reciprocal basis.
 \label{coordinate} }
\end{figure}

To simplify the problem, we describe the magnetic ordering by a
version of mean-field theory, in which one writes the magnetic free
energy \cite{Harris2007} $F_M$ as

\begin{eqnarray}
H=\frac{1}{2}\sum_{\textbf{r}_i,\alpha;\textbf{r}_j,\beta}
\chi^{-1}_{\alpha\beta}(\textbf{r}_i,\textbf{r}_j)
S_\alpha(\textbf{r}_i)S_\beta(\textbf{r}_j) + O(S^4),
\end{eqnarray}
where $S_{\alpha}(\textbf{r}_i)$ is the thermally averaged $\alpha$
component of the spin at position $\textbf{r}_i$. Introducing
Fourier transformations of $S_\alpha (\textbf{r}_i)$ and
$\chi_{\alpha\beta}^{-1}(\textbf{r}_i, \textbf{r}_j)$ and omitting
high order terms, we have

\begin{eqnarray}
&&F_M =
\frac{1}{2}\sum_{\textbf{q};\tau_i,\tau_j,\alpha\beta}\chi_{\alpha\beta}^{-1}(\textbf{q};\tau_i\tau_j)
S_\alpha(-\textbf{q};\tau_i)S_\beta(\textbf{q};\tau_j),\\
&&S_\alpha(\textbf{q},\tau_i) =
\frac{1}{N}\sum_{\textbf{R}}S_\alpha(\textbf{R}+\tau_i)e^{i\textbf{q}\cdot(\textbf{R}+\tau_i)},\\
&&\chi_{\alpha\beta}^{-1}(\textbf{q};\tau_i,\tau_j)=
\sum_{\textbf{R}}\chi_{\alpha\beta}^{-1}(\tau_i,\textbf{R}+\tau_j)e^{i\textbf{q}\cdot(\textbf{R}+\tau_j-\tau_i)},
\end{eqnarray}
where N is the number of the unit cells in the system, $\tau_i$ is
the location of the $i$th site within the unit cell ($\tau_i$ is
(0,0,0) in MnI$_2$), and $\textbf{R}$ is the lattice vector. As our
main interest lies in explaining the observed magnetic modulation
vector $\textbf{q}$, we have completely ignored anisotropy, whose
major effect is to select the spin orientations. So we have an
isotropic model,
\begin{eqnarray}
\chi_{\alpha\beta}^{-1}=J_{\alpha\beta}(\tau_i,\tau_j)\delta_{\alpha\beta}+[K+dk_B
T]\delta_{\alpha\beta}\delta_{\tau_i\tau_j},
\end{eqnarray}
where $\delta_{\alpha\beta}$ is unity if $\alpha=\beta$ and is zero
otherwise. $d$ is a spin-dependent constant of order unity, so that
$-dk_B\sum_\alpha S_\alpha(\textbf{r})^2$ is the entropy (relative
to infinite temperature) associated with a spin $S$. In our case, we
only consider the exchange couplings defined in
Fig.~\ref{exchange_parameter}. In our coordinate system, the lattice
vectors are $\vec{a}_1=a\vec{i}$,
$\vec{a}_2=-\frac{1}{2}a\vec{i}+\frac{\sqrt{3}}{2}a\vec{j}$ and
$\vec{a}_3=c\vec{k}$ (see Fig.~\ref{coordinate}, x1, y1), where $a$
and $c$ are the lattice constants of MnI$_2$. The reciprocal vectors
are $\vec{b}_1=\frac{2\pi}{a}(\vec{i}+\frac{\sqrt{3}}{3}\vec{j})$,
$\vec{b}_2=\frac{2\pi}{a}\cdot\frac{2\sqrt{3}}{3}\vec{j}$ and
$\vec{b}_3=\frac{2\pi}{c}\vec{k}$. Setting
$\textbf{q}=q_1\vec{b}_1+q_2\vec{b}_2+q_3\vec{b}_3$ in the Fourier
transformation, we have the following $\chi^{-1}(\textbf{q})$ with
some algebra in MnI$_2$,

\begin{eqnarray}  \label{Fourier}
\chi^{-1}(\textbf{q}) &=&K+ d k_B T \nonumber \\
&+& 2J_1[\cos(q_1)+\cos(q_1+q_2)+\cos(q_2)]\nonumber \\
&+&2J_{ab}[\cos(2q_1+q_2)+\cos(q_1+2q_2)\nonumber \\
&+&\cos(-q_1+q_2)]\nonumber \\
&+&2J_2[\cos(2q_1)+\cos(2q_1+2q_2)+\cos(2q_2)]\nonumber \\
&+&4J_{NC}[\cos(q_1)+\cos(q_1+q_2)+\cos(q_2)]\cos(q_3)\nonumber \\
&+&2J_C\cos(q_3)+2J_{NNC}[\cos(q_1+2q_2+q_3)\nonumber \\
&+&\cos(2q_1+q_2-q_3)+\cos(q_1-q_2+q_3)].
\end{eqnarray}

\begin{figure}
\centerline{\includegraphics[height=4.0cm]{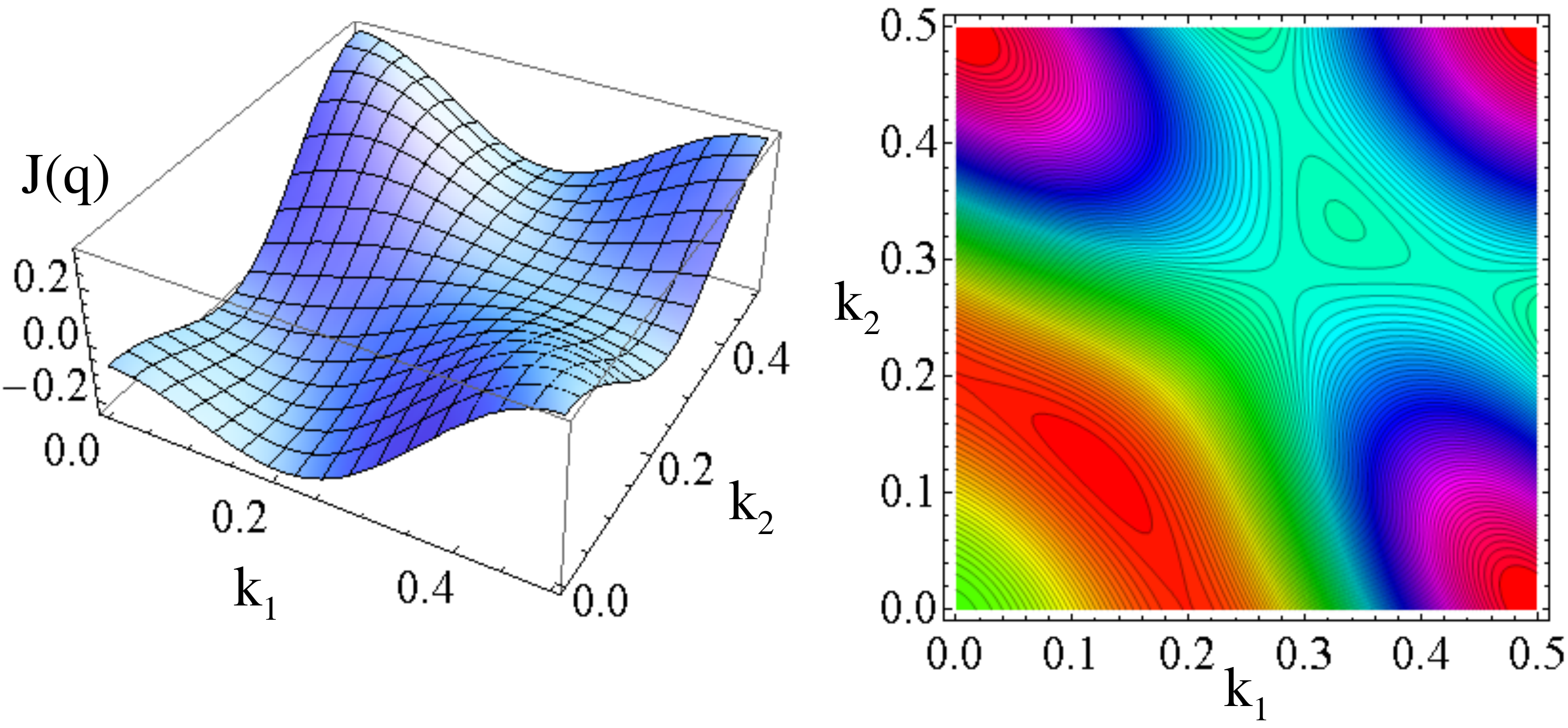}}
\caption{(color online). The diagram of $\chi^{-1}(q_1,q_2,q_3)$
with $q_3$ fixed as experimental value 0.5 at $T_{N1}$. The minimum
point is at $(q_1, q_2)=(0.1226,0.1226)$, which is close to the
experiment value $(0.1025, 0.1025)$.
 \label{z05} }
\end{figure}

\begin{figure}
\centerline{\includegraphics[height=4.0cm]{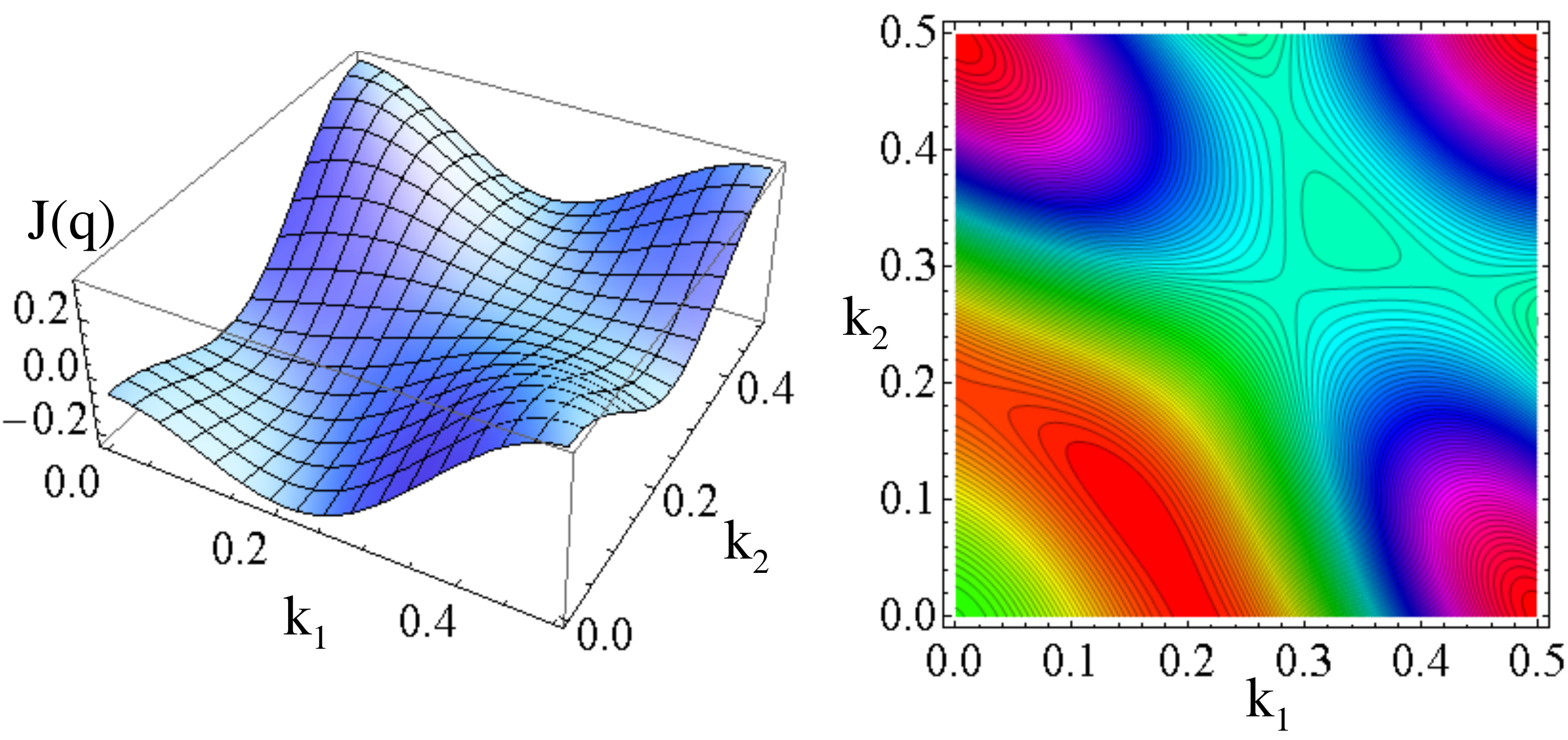} }
\caption{(color online). The diagram of $\chi^{-1}(q_1,q_2,q_3)$
with $q_3$ fixed as experimental value 0.439 at $T_{N3}$. The
minimum point is at $(q_1, q_2)=(0.155,0.089)$, which is consistent
with the experiment value $(0.181, 0)$.
 \label{z044} }
\end{figure}

Setting $\chi^{-1}(\textbf{q})= K+ d k T+J(\textbf{q})$ and
substituting the exchange parameters calculated with $U_{eff}=4$ eV
in Table \ref{exchange} into Eq.~(\ref{Fourier}), one can easily
obtain the free energy surface in $\textbf{q}$($q_1$, $q_2$, $q_3$)
space. The minimum points of this surface might correspond to the
experimentally determined magnetic modulation vectors at different
temperature. In Fig.~\ref{z05}, by fixing $q_3$=0.5, the minimum
point of $\chi^{-1}$ is at $(q_1, q_2)=(0.1226,0.1226)$, which is in
good agreement with experimental value $(0.1025, 0.1025)$ at
transition temperature $T_{N1}$. By setting $q_3$=0.439, as shown in
Fig.~\ref{z044}, we get $(q_1, q_2)=(0.155,0.089)$ after minimizing
$\chi^{-1}$, which is also consistent with the experimental value
$(0.181, 0)$ at $T_{N3}$. Assuming $q_2$=0 in the ground state, we
minimize $\chi^{-1}(q_1,q_3)$ and obtain $(q_1, q_3)=(0.206,0.444)$,
which also reproduces the experimental values (0.181,0.439) at
$T_{N3}$. Therefore, we believe $U_{eff}$=4 eV is proper for Mn in
MnI$_2$ for the GGA+U calculation.

It is of interest to notice that $\textbf{q}$ vector has
nonequivalent $q_1$ and $q_2$ when temperature is below $T_{N3}$,
which means that $\chi^{-1}(\textbf{q})$ should have asymmetric
terms when exchanging $q_1$ and $q_2$. The only possible term is
that determined by $J_{NNC}$ coupling with proper choice of $q_3$.
When the value $q_3$ is 0 or 0.5, $J(\textbf{q})$ is invariant under
the exchange of $q_1$ and $q_2$. However, if $q_3$ is neither 0 nor
0.5, from Eq.~\ref{Fourier} we find that the term contributed by
$J_{NNC}$ makes $q_1$ and $q_2$ inequivalent. Thus $J_{NNC}$ is of
crucial importance to the magnetic ground state, where $q_1$ is not
equal to $q_2$. Although the two layers of MnI$_2$ have a large
separation about 3.5 \AA, it cannot be treated as a quasi-2-D
triangle lattice of Mn atoms due to the important interplane
coupling $J_{NNC}$.

\begin{figure}
\centerline{\includegraphics[height=4.0cm]{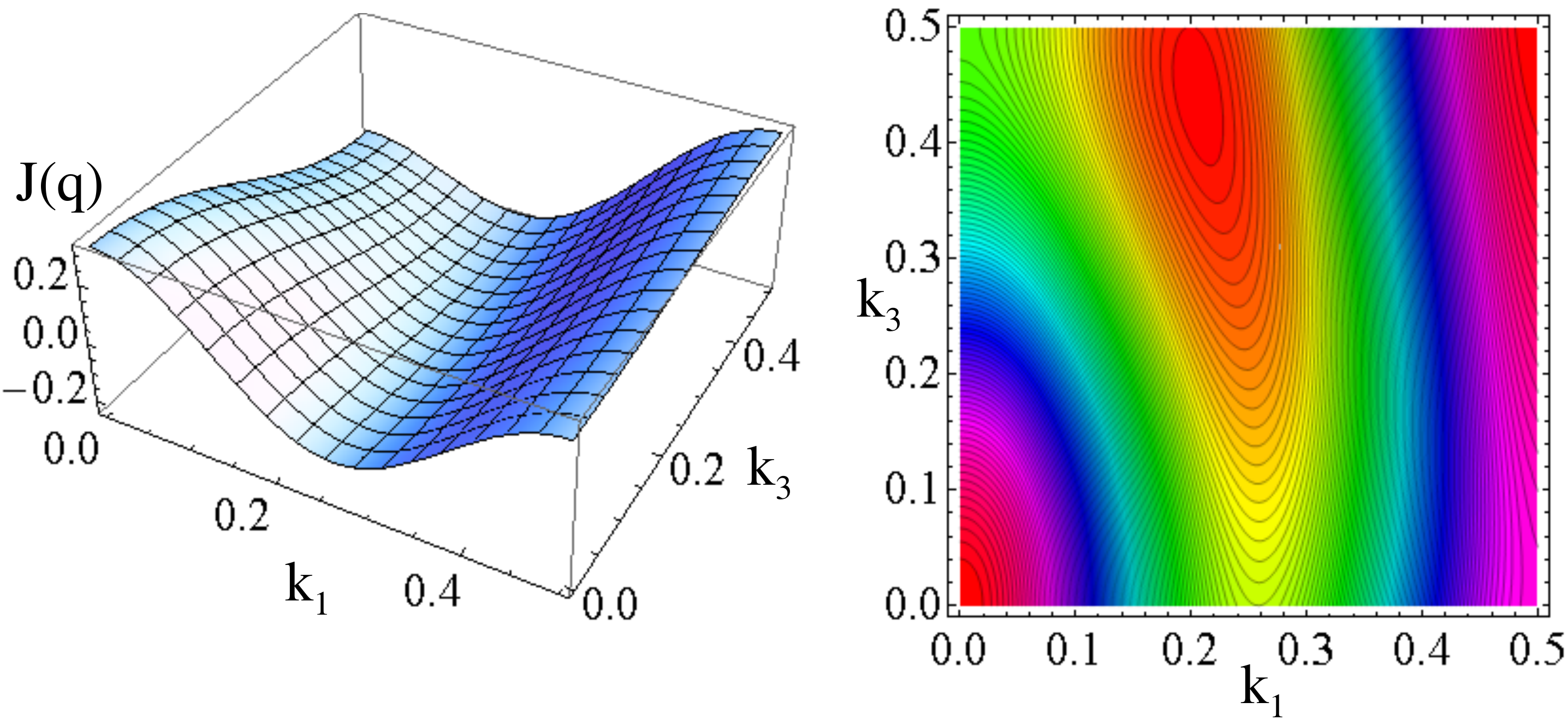}}
 \caption{(color online). The diagram of $\chi^{-1}(q_1,q_2,q_3)$ with $q_2$ fixed as experimental value 0 at $T_{N3}$. It is noted from above picture the minimum point is at
$(q_1, q_3)=(0.206,0.444)$, which is close to the experiment value
$(0.181, 0.439)$.
 \label{kxkz} }
\end{figure}

In the mean-field approximation\cite{Smart1966,Nagamiya1967}, the
transition temperature $T_N$ and Curie-Weiss temperature
$\theta_{CW}$ are related to spin exchange parameters as

\begin{eqnarray}
T_{N}=-\frac{S(S+1)}{3k_B}J(\textbf{q})_{min},
\\
\theta_{CW}=-\frac{S(S+1)}{3k_B}\sum_i Z_iJ_i,
\end{eqnarray}
where the sum is over all the nearest neighbors of a given spin
site, $Z_i$ is the number of the nearest neighbors connected by
exchange coupling $J_i$, and $S$ is the spin quantum number on each
site(in present case $S=5/2$). Therefore, the $T_N$ and
$\theta_{CW}$ are estimated to be 8.7 K and -10.7 K, respectively,
using exchange parameters calculated from $U_{eff}$=4 eV, which is
comparable with the experimental $T_N$ 3.95 K and $\theta_{CW}$ -8
K\cite{Hoekstra}. The ratio of the Curie-Weiss and the magnetic
ordering temperature $\alpha \equiv|\theta_{CW}/T_{N1}|$ has been
proposed as a quantitative measurement of frustration. In MnI$_2$
the ratio is about 2 in experiment, which is comparable with that of
RbFe(MoO$_4$)$_2$($\alpha \sim6$)\cite{Gasparovic} and is quite a
low value with respect to NaTiO$_2$($\alpha
>500$)\cite{Hirakawa}.

To end this section, we briefly give a summary. We obtain the
exchange parameters through DFT calculation and use these parameters
in Heisenberg exchange model to calculate the magnetic modulation
vectors at different magnetic phase of MnI$_2$. The intraplane
couplings lead to frustration in the triangle lattice of MnI$_2$
while the interplane coupling also has an important contribution.
The nearest and next-nearest coupling compete with each other, which
makes the spiral magnetism stable. The magnetic vectors obtained are
in good agreement with those in experiment. The choice of $U_{eff}$
is very important. In general, the criterion for choosing U value is
to see whether it can well reproduce the experimental measurements
systematically. In our case, we have checked that the calculated
magnetic moment and the resulted magnetic modulation vectors,
transition temperature, Currie-Weiss temperature, as well as the
ferroelectric polarization are in good agreement with the
experimental measurements when $U_{eff}$=4.0 eV. While other
$U_{eff}$  values (3, 5 and 6 eV)  lead to large deviation or even
qualitative error in some, or all of these physical quantities.

\section{ferroelectricity of $\textrm{MnI}_2$}\label{ferroelectricity}\label{S2}

\subsection{Symmetry analysis of MnI$_2$}
In this section, we perform a group theoretical calculation for
MnI$_2$ magnetic structure and determine the direction of the FE
polarization. The ferroelectric polarization appears when
temperature is below 3.45 K with a magnetic vector q$\sim$(0.181, 0,
0.439)\cite{Kurumaji}. The general positions of ions with space
group $P\bar{3}m1$ are given in Table~\ref{position}.

\begin{table}[t]
\caption{\label{position}%
General positions for $P\bar{3}m1$. Here "3" and "2" denote three
fold and two fold rotation, respectively. $m_n$ label the three
mirror planes.}
\begin{ruledtabular}
\begin{tabular}{lll}
 $E\textbf{r}=(x,y,z)$ & $3\textbf{r}=(\bar{y},x+\bar{y},z)$ & $3^2
\textbf{r}=(\bar{x}+y,\bar{x},z)$
\\
$2_1\textbf{r}=(y,x,\bar{z})$ &
$2_2\textbf{r}=(x+\bar{y},\bar{y},\bar{z})$ &
$2_3\textbf{r}=(\bar{x},\bar{x}+y,\bar{z})$
\\
$I\textbf{r}=(\bar{x},\bar{y},\bar{z})$ &
$I3\textbf{r}=(y,\bar{x}+y,\bar{z})$ &
$I3^2\textbf{r}=(x+\bar{y},x,\bar{z})$
\\

$m_1\textbf{r}=(\bar{y},\bar{x},z)$ &
$m_2\textbf{r}=(\bar{x}+y,y,z)$ & $m_3\textbf{r}=(x,x+\bar{y},z)$

\end{tabular}
\end{ruledtabular}
\end{table}

Considering the wave vector $\textbf{q}=q_x\hat{i}+q_z\hat{k}$ (in
our second Cartesian coordinate x2, y2, see Fig.~\ref{coordinate} ),
it is clear that the only operation (other than the identity) which
conserves the wave vector is $m_3$ (mirror plane with respect to xz
plane). We adopt the method in Ref.~\onlinecite{Harris2007} to
analyze the polarization of MnI$_2$. Clearly, the Fourier component
$S_x(\textbf{q})$ obeys
\begin{eqnarray}
m_3S_x(\textbf{q})&=&\lambda(m_3)S_x(\textbf{q}),  \nonumber \\
m_3S_z(\textbf{q})&=&\lambda(m_3)S_z(\textbf{q}),
\end{eqnarray}
with $\lambda(m_3)=-1$, and that is irrep $\Gamma_1$. For irrep
$\Gamma_2$, $\lambda(m_3)=1$ and we have
\begin{eqnarray}
m_3S_y(\textbf{q})&=&\lambda(m_3)S_y(\textbf{q}).
\end{eqnarray}

To fix the complex coefficients, we consider the effect of
inversion, which leads to
\begin{eqnarray}\label{inversion}
IS_\alpha(\textbf{q})&=&S_\alpha^{*}(\textbf{q}).
\end{eqnarray}
For $\Gamma_1$, we consider its quadratic free energy and substitute
Eq.~(\ref{inversion}), then $S_x(\textbf{q})$ and $S_z(\textbf{q})$
will have the same complex phase.\cite{Harris2007} We now introduce
order parameters which describe the magnitude and phase of these two
symmetry labels (irreps). When both irreps are present, one has
\begin{eqnarray}
S_x(\textbf{q})&=&\bm{\sigma }_1(\textbf{q})r,  \nonumber \\
S_z(\textbf{q})&=&\bm{\sigma }_1(\textbf{q})s,  \nonumber \\
S_y(\textbf{q})&=&\bm{\sigma }_2(\textbf{q}),
\end{eqnarray}
where $r^2+s^2=1$ (r and s are real) and $\textbf{$ \sigma
$}_n(\pm\textbf{q})=\sigma_n e^{\mp i\phi_n}$. We also have the
transformation properties
\begin{eqnarray}
m_3\bm{\sigma }_1=-\bm{\sigma }_1, m_3\bm{\sigma }_2=\bm{ \sigma }_2,  \nonumber \\
I\bm{ \sigma }_1=\bm{ \sigma }_1^{*}, I\bm{ \sigma }_2=\bm{\sigma
}_2^{*}.
\end{eqnarray}
When both irreps are present, we have
\begin{eqnarray}
S_x(\textbf{r})&=&\sigma_1(\textbf{q})r\cos(\vec{q}\cdot\vec{r}+\phi_1),  \nonumber \\
S_y(\textbf{r})&=&\sigma_2(\textbf{q})\cos(\vec{q}\cdot\vec{r}+\phi_2),  \nonumber \\
S_z(\textbf{r})&=&\sigma_1(\textbf{q})s\cos(\vec{q}\cdot\vec{r}+\phi_1).
\end{eqnarray}

Now we consider the magnetoelectric coupling in MnI$_2$ using order
parameter obtained above. Since single order parameter cannot
produce ferroelectricity in our case, we consider both irreps,
\begin{eqnarray}
F_{int}=i\sum_{\gamma}r_{\gamma}P_\gamma[\bm{
\sigma}_1(\textbf{q})\bm{\sigma}_2^{*}(\textbf{q})-\bm{\sigma
}_1^{*}(\textbf{q})\bm{\sigma}_2(\textbf{q})].
\end{eqnarray}

Under operation $m_3$, $\bm{\sigma}_1(\textbf{q})\bm{ \sigma
}_2^{*}(\textbf{q})$ and $\bm{\sigma}_1^{*}(\textbf{q})\bm{\sigma
}_2(\textbf{q})$ will change sign. Since that $F_{int}$ is invariant
under $m_3$, it requires $P_\gamma$ to be odd under $m_3$, so
$\vec{P}$ has to be along $\hat{y}$ direction (y2 direction in our
second Cartesian coordinate), which is found in
experiment\cite{Kurumaji}.

\subsection{Calculating the polarization using DFT}

\begin{figure*}
\centerline{\includegraphics[height=8.0cm]{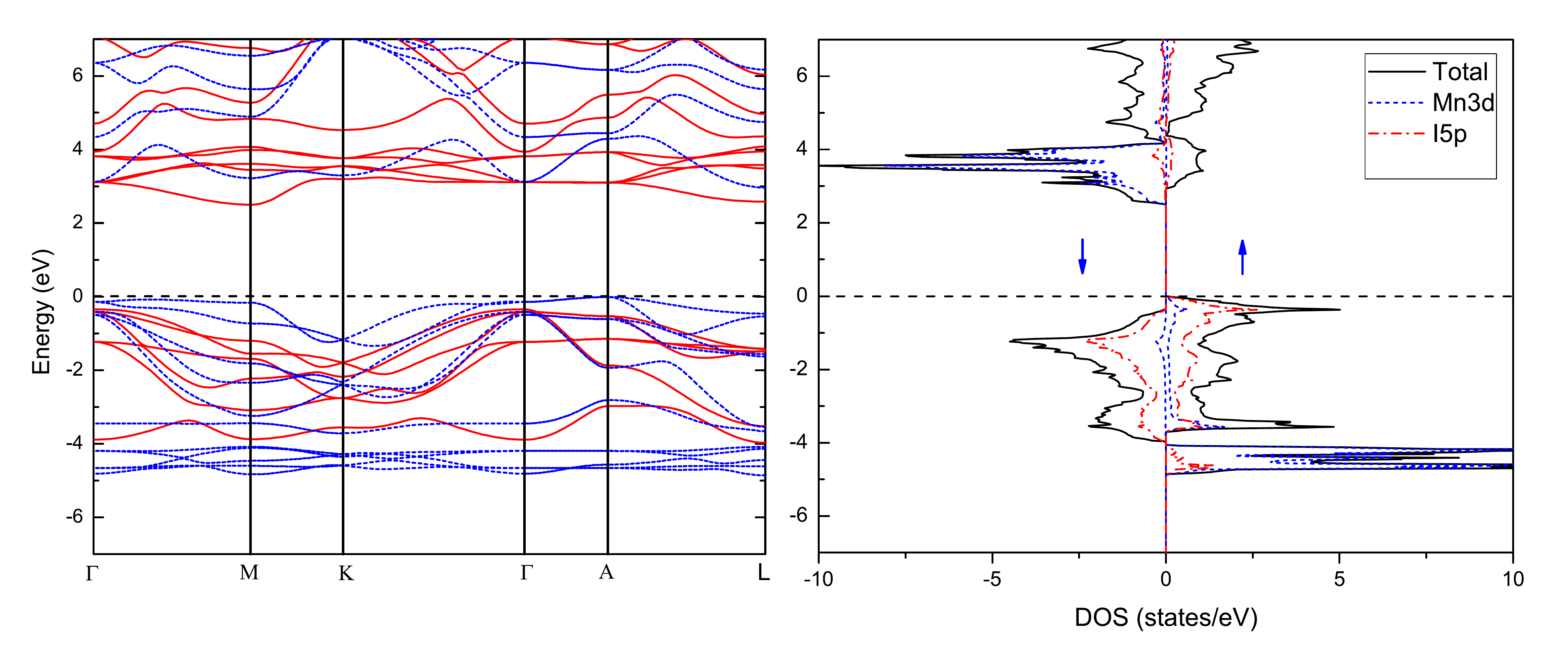}}
\caption{(color online). Band structure and density of states(DOS)
for MnI$_2$ calculated with GGA+U ($U_{eff}$=4 eV). Left panel is
the band structure plot, the blue dash lines and red solid lines
denote spin-up band and spin-down band, respectively. $\Gamma$ is
the center of the Brillouin zone , M (0,0.5,0),
K($\frac{1}{3}$,$\frac{1}{3}$,0), A (0,0,0.5) and L (0,0.5,0.5) in
reciprocal lattice. Right panel is the DOS plot, the positive and
negative value of DOS denote the majority spin states and the
minority spin states, respectively. \label{band_dos_FM} }
\end{figure*}

The electronic structure of MnI$_2$ calculated for FM state with
$U_{eff}=4$ eV is presented in Fig.~\ref{band_dos_FM}. It is clear
that FM state is insulating with an indirect band gap. Mn 3d states
are mainly located in the lower energy region from -5.0 eV to -4.0
eV in the spin-up channel, and they are almost empty for the
spin-down channel. Therefore, the Mn$^{2+}$ ions in MnI$_2$ are high
spin. The narrow and high peaks in density of states plot indicate
that the 3d electrons of Mn are localized. The top of the valence
band is primarily attributed to I 5p states, hybridized weakly with
Mn 3d. The bottom of conduction band is mainly attributed to Mn 3d
down spin states. The band dispersion is strong in $ab$ plane but
weak in $c$ direction near fermi energy, as expected for the layered
structure of MnI$_2$.

As it has been shown that the propagation vector of MnI$_2$ is
$\textbf{q}=(0.181, 0, 0.439)$, we perform GGA+U with the
experimental $\textbf{q}$ in absence of SOC, in which just one
primitive cell is used due to the generalized Bloch
theorem\cite{Sandratskii}. Then we calculate the electric
polarization using the Berry phase method\cite{King-Smith}. However,
negligible polarization is found. The above observation leads us to
consider SOC effect on the electric polarization in the spiral state
of MnI$_2$. We carry out GGA+U+SOC ($U_{eff}=4$ eV) calculation for
the $\textbf{q}=(0.25, 0, 0.5)$ \cite{vector_comment} spiral states
with spin in (307) plane\cite{Cable}. The polarization is 107.3 $\mu
C/m^2$ along a2 direction, which is consistent with polarization
$\textbf{P}\perp\textbf{q}_{in}$ (0.25, 0, 0) \cite{Kurumaji}. The
experimental polarization is 84 $\mu C/m^2$ at 2 K and the
interpolated value at 0 K is about 128 $\mu C/m^2$. The calculated
polarization is a little bit smaller than the interpolated value,
which is due to the approximation of the magnetic vector
$\textbf{q}$. As depicted in Ref.~\onlinecite{Kurumaji}, the
magnetic vector is parallel to [100] (in our coordinate is along OD,
see Fig.~\ref{coordinate}), when high magnetic field along [100] is
applied. In this case, we perform calculation with
$\textbf{q}=(\frac{1}{3}, \frac{1}{3}, 0)$ and the ferroelectric
polarization is about 58 $\mu C/m^2$ along OD, which is very close
to 57 $\mu C/m^2$ at $\theta_H=30^ \circ$ in high magnetic
field\cite{Kurumaji} and consistent with experiment
$\textbf{P}\parallel\textbf{q}_{in}$ $(\frac{1}{3}, \frac{1}{3}, 0)$
but not with the prediction from KNB model\cite{Katsura2005}.

\begin{figure}
\centerline{\includegraphics[height=12.0
cm]{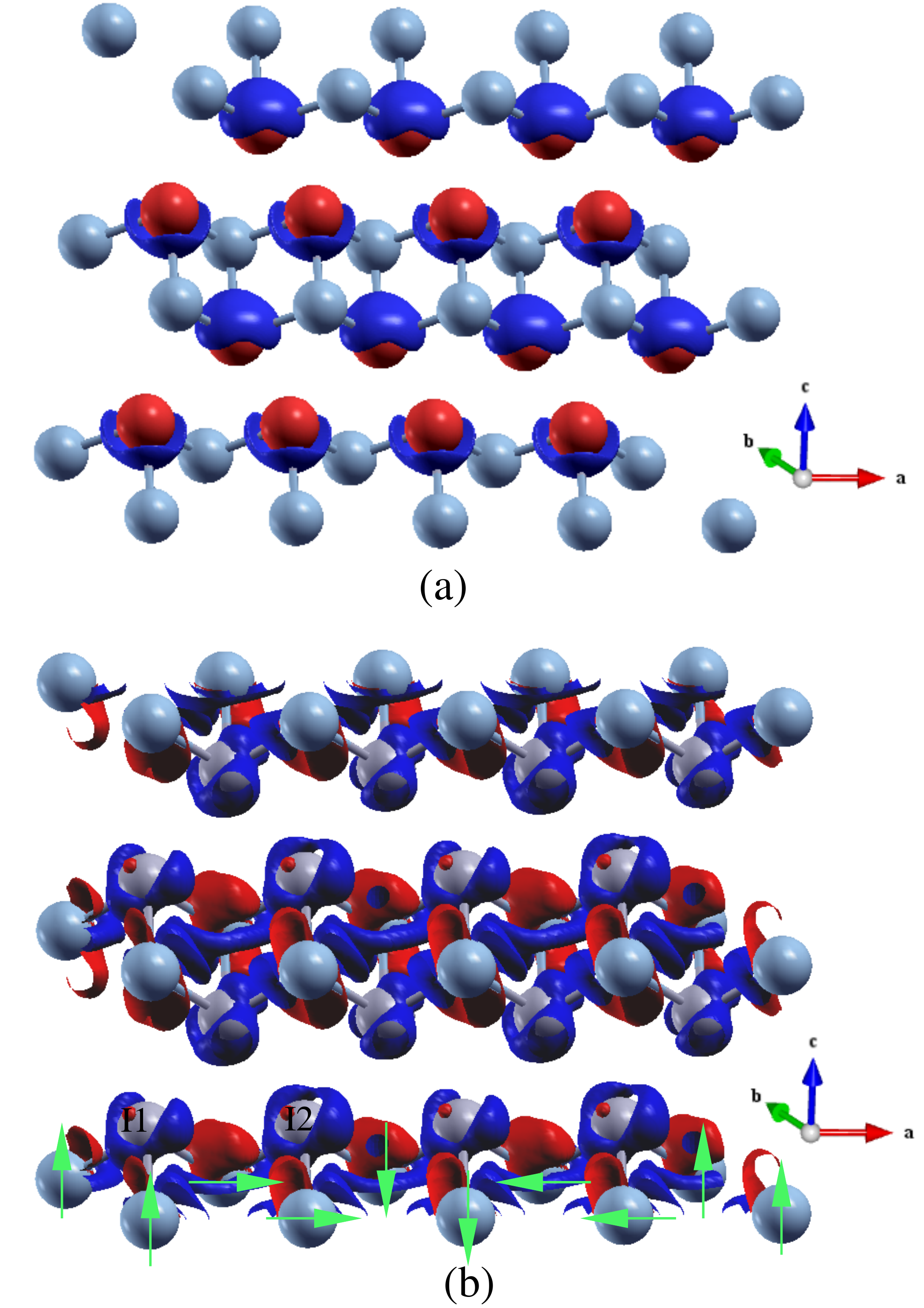}}
 \caption{(color online).
(a) Perspective view of an isosurface calculated for the electron
density difference between the GGA+U+SOC and GGA+U results for the
case of \textbf{q}=(0.25, 0, 0.5). The red and blue surface
represent 5.5$\times10^{-4} e/$\AA$^{3}$ and -5.5$\times10^{-4}
e/$\AA$^{3}$, respectively. (b) Perspective view of an isosurface
calculated for the electron density difference between the spiral
magnetic state (\textbf{q}=(0.25, 0, 0.5)) and FM state with
GGA+U+SOC. The red and blue surface represent 5.0$\times10^{-5}
e/$\AA$^{3}$ and -5.0$\times10^{-5} e/$\AA$^{3}$, respectively. The
arrows on Mn atoms denote the direction of their spins. The electron
distribution on I1 and I2 has visible difference.
 \label{charge_diff} }
\end{figure}

Obviously, the polarization of MnI$_2$ is induced by SOC, which is
also consistent with the calculation of Xiang \emph{et
al}.\cite{Xiang2011}. 
The spin-orbit part of Hamiltonian is $\hat{H}_{SO}=\lambda
\hat{\bm{L}}\cdot \hat{\bm{S}}$, where $\lambda$ is the spin-orbit
coupling constant. SOC is expected to be strong on I 5p orbitals as
the $\lambda$ increases with the nuclear charge of the atom and
decreases with increasing quantum numbers (angular quantum number
and principle quantum number).

To examine how the polarization arises from the spiral magnetic
state with SOC, we analyze the electron distribution of MnI$_2$ by
showing the difference in electron density between calculations with
and without SOC included for the case of $\textbf{q}=(0.25, 0,
0.5)$. As shown in Fig.~\ref{charge_diff}(a) that the main
asymmetric charge distribution is around each I ion, which makes a
primary contribution to the ferroelectric polarization. The
graphical software XCrysDen\cite{XCrysDen} was used to plot the
electron density difference. The tremendous changes of charge
density around I indicate that the SOC on I is rather strong. To
study how spiral magnetism contributes to FE polarization, we
analyze the electron distribution by showing the difference between
the electron density of spiral state with $\textbf{q}=(0.25, 0,
0.5)$ and that of FM state (a rather good reference state) of
MnI$_2$. Both calculations are performed with GGA+U+SOC. There is no
FE polarization in FM state because the inversion symmetry is
preserved, while it is broken by spiral magnetism, which is
essential to appearance of FE polarization in MnI$_2$. From
Fig.~\ref{charge_diff}(b), we find that there are changes of
electron density around both Mn and I atoms. The strength of SOC on
Mn is weaker compared with that of I, but it is also important.
Spin-orbit interaction couples spin moment with electron's spacial
orbital. As a result, the changes of spin direction will influence
the spacial distribution of charge. The strong hybridization of Mn
3d and I 5p, as well as the strong SOC on I, will result in the
asymmetric charge distribution around I atoms depending on the spin
states of surrounding Mn atoms. That is why the electron density
around I1 and I2(in Fig.~\ref{charge_diff}(b)) looks different.
Therefore, both spiral magnetism and SOC are essential to the FE
polarization of MnI$_2$. The spiral magnetism breaks the inversion
symmetry and the degree of freedom of orbital couples with that of
spin through SOC, which lead to asymmetric charge distribution (this
is FE polarization). For the case of
$\textbf{q}=(\frac{1}{3},\frac{1}{3},0)$, the electron density
difference is similar to the case of $\textbf{q}=(0.25, 0, 0.5)$.
Furthermore, we find a linear relationship between the magnitude of
electric polarization and the strength of SOC.

 To examine the effect of ion
displacement in the spin-spiral state on the FE polarization, we
optimize the atoms positions of MnI$_2$ in the above two cases by
performing GGA+U+SOC calculation until the atomic forces become less
than 0.02 $eV/$\AA$ $ and then calculate the electric polarization
using the relaxed structures. In the case of $\bm{q}=(0.25,0,0.5)$,
it is found that Mn$^{2+}$ ions move along a2 direction, which leads
to slightly enhanced FE polarization to about 170 $\mu C/m^2$ in
comparison with the value of 107.3 $\mu C/m^2$ without ion
displacement. For $\bm{q}=(\frac{1}{3},\frac{1}{3},0)$, it is found
that the sum of Mn displacements is along OD direction while that of
I ions is along the opposite direction.  The in-plane electric
polarization of the relaxed structure is greatly enhanced from 58
$\mu C/m^2$ to 170 $\mu C/m^2$ with direction of FE polarization
reversed. The out-of-plane component in this case is about one third
of the in-plane component.

\section{summary and conclusions}\label{S3}

In this paper we have presented a comprehensive investigation of the
incommensurate magnetic structure and ferroelectric polarization of
the new multiferroic material MnI$_2$. Six exchange interaction
parameters among local moments on Mn sites are obtained by mapping
the mean-filed Heisenberg model Hamiltonian onto the total energy
differences of eight different magnetic ordering states from DFT
calculations. We show the inter-plane coupling $J_{NNC}$ is fairly
strong because of its linear exchange path, as suggested before by
Wollan \emph{et al}.\cite{Wollan}, although there is a large
separation between the Mn-layers. As a result, the lattice of Mn
cannot be simply treated as a quasi two dimensional system.
Moreover, this inter-plane coupling $J_{NNC}$ is critical to
generating spiral spin structure by breaking the equivalence of the
spin-density wavevectors along two different directions, $q_1$ and
$q_2$, in the magnetic ground state. Our calculation also indicates
that Hubbard $U_{eff}$ strongly affects the magnetic exchange
couplings. It is found that $U_{eff}\sim4$ eV can give results
quantitatively consistent with experimental values. For example, the
Currier-Weiss temperature is estimated as -10.7 K when $U_{eff}=4$
eV, which is very close to -8 K in the experimental
measurement\cite{Hoekstra}.

We also use both symmetry analysis and DFT  calculations to investigate the
polarization of this material. Our study reveals that SOC is
essential for its ferroelectric polarization. Both the direction and
magnitude of the polarization obtained from DFT calculation are in
good agreement with the experimental data. Charge density difference
analysis shows that the primary asymmetric charge distribution is
around I ions, due to their strong SOC effect.

The isotropic Heisenberg model considered in this paper provides a
good description of the magnetic ordering in MnI$_2$.  This suggests
that the polarization induced by spiral magnetic ordering has no
strong feedback effect on magnetic ordering.  This result is
consistent with the observed small value of polarization.

\section{Acknowledgments}

 We thank  J.M. Zhang, J. Ding, Z. Liu and Y. R. Zhang for extremely useful
 discussion.  The work is supported by "973" program (Grant No.
 2010CB922904 and No. 2012CV821400), as well as  national science foundation of China ( Grant No. NSFC-1190024, 11175248
and 11104339).

\end{document}